\documentclass[11pt]{article}

\usepackage{epsfig}

\def\al{\alpha^{\prime}}
\def\vp{\varphi} 
\def\rd{{\rm d}}

\def\bea{\begin{eqnarray}}
\def\eea{\end{eqnarray}}
\def\beq{\begin{equation}}
\def\eeq{\end{equation}}

\def\be{\begin{equation}}
\def\ee{\end{equation}}

\def\5{\overline 5}
\newcommand{\gsim}{\mbox{\raisebox{-1.ex}{$\stackrel
     {\textstyle>}{\textstyle\sim}$}}}

\newcommand{\square}{\kern1pt\vbox{\hrule height
1.2pt\hbox{\vrule width 1.2pt\hskip 3pt
   \vbox{\vskip 6pt}\hskip 3pt\vrule width 0.6pt}\hrule
height 0.6pt}\kern1pt}
\begin{document}

\begin{center}

\thispagestyle{empty}

\begin{center}

\vspace{1.7cm}

{\LARGE\bf Construction of nonsingular cosmological solutions \\
in string theories}

\end{center}

\vspace{1.4cm}

{\Large Shinji Tsujikawa}

\vspace{1.0cm}
{\em Research Center for the Early Universe, University of Tokyo,} \\
{\em Hongo, Bunkyo-ku, Tokyo 113-0033, Japan}\\
\vspace{0.3cm}
{\em email: shinji@resceu.s.u-tokyo.ac.jp\\}

\vspace{.3cm}

\end{center}

\vspace{3cm}

\centerline{\Large Abstract}
\vspace{3 mm}
\begin{quote}\small
We study nonsingular cosmological scenarios in a general $D$-dimensional 
string effective action of the dilaton-modulus-axion system in the presence 
of the matter source.  In the standard dilatonic Brans-Dicke parameter 
($\omega=-1$) with radiation, we analytically obtain singularity-free 
bouncing solutions where the universe starts out in a state with a finite 
curvature and evolves toward the weakly coupled regime.  We apply this 
analytic method to the string-gas cosmology including the massive state in 
addition to the leading massless state (radiation), with and without 
the axion.  We numerically find bouncing solutions which asymptotically 
approach an almost radiation-dominant universe with a decreasing curvature 
irrespective of the presence of the axion, implying that inclusion of the 
matter source is crucial for the existence of such solutions for 
$\omega=-1$.  In the theories with $\omega \ne -1$, it is possible to 
obtain complete regular bouncing solutions with a finite dilaton and 
curvature in both past and future asymptotics for the general dimension, 
$D$.  We also discuss the case where dilatonic higher-order corrections are 
involved to the tree-level effective action and demonstrate that the 
presence of axion/modulus fields and the matter source does not 
significantly affect the dynamics of the dilaton-driven inflation and the 
subsequent graceful exit.
\end{quote}

\vfill 

\newpage

\section{Introduction}                            

String cosmology has continuously received much attention 
\cite{Lidsey:1999mc} with the development of string theory as a possible 
candidate to unify all fundamental interactions in nature \cite{str}.  
For cosmologists it is very important to test the viability of string 
theory by extracting cosmological implications from it.  In particular 
string cosmology has an exciting possibility to resolve the big-bang 
singularity which plagues in General Relativity.

Among the string-inspired cosmological scenarios proposed so far, 
the pre-big-bang (PBB) model \cite{Veneziano:1991ek,pbb} 
based on the low energy, tree-level string effective action has been 
widely studied (see ref.~\cite{Gasperini:2002} for a recent review).  In 
this scenario there exist two disconnected branches, one of which corresponds 
to the dilaton-driven inflationary stage and another of which is the 
Friedmann branch with decreasing curvature.  Then string corrections to the 
effective action can be important around the high curvature regime where 
the branch change occurs.  Recently proposed Ekpyrotic/Cyclic cosmologies 
\cite{ekpyr,cyclic} have a similarity to the PBB scenario in the sense that 
the description in terms of the tree-level effective action breaks down around 
the collision of two branes in a five dimensional bulk.

When the universe evolves toward the strongly coupled, high curvature regime 
with growing dilaton, it is inevitable to implement higher-order corrections 
to the tree-level action.  Indeed it was found that two branches can be 
smoothly joined by taking into account the dilatonic higher-order derivative 
and loop corrections in the context of PBB \cite{Gasperini:1996fu,Bru} and 
Ekpyrotic \cite{Tsujikawa:2002qc} scenarios.  In the system where a modulus 
field is dynamically important rather than the dilaton, 
Antoniadis {\it et al.} showed that the big-bang singularity can be avoided 
by including the Gauss-Bonnet curvature invariant coupled to the modulus 
\cite{Antoniadis:1993jc}.

In contrast we can consider the case where the dilaton 
evolves toward the weakly coupled region.
In the brane-gas cosmology \cite{Alex,Brand,Easson,EGJ,Easther} and the 
recently proposed string-gas cosmology \cite{Bassett:2003ck}, the universe 
starts out from a dense and hot state filled with a gas of fundamental 
branes and strings in thermal equilibrium.  These are originated from the 
work of Brandenberger and Vafa \cite{BV} (see also ref.~\cite{TV}) in  an 
attempt to construct singularity-free cosmological scenarios using 
T-duality.  In these models it is typically assumed that the cosmic 
evolution does not reach the strongly coupled regime in order to avoid the 
breakdown of the effective ten-dimensional background description (see 
ref.~\cite{Maggiore:1998cz} for the parameter regions where the 
effective ten-dimensional description of the physics is no longer valid).  In 
particular, in the context of the string-gas cosmology, it was shown that only 
the large 3 dimensions expand and asymptotically approach the 
radiation dominant universe while the small 6 or 7 dimensions are kept 
undetectably small \cite{Bassett:2003ck}.

In order to obtain viable nonsingular cosmological scenarios in string 
theories, the string coupling and the scalar curvature are required to be 
finite during the cosmological evolution.  In the standard tree-level 
action with the dilatonic Brans-Dicke parameter $\omega=-1$, it is generally 
difficult to construct nonsingular solutions in the absence of the matter 
source term.  This comes from the fact that the evolution toward the 
strongly coupled regime reaches the curvature singularity in the tree-level 
action and that the evolution toward the weakly coupled regime does not 
have enough power to lead to bouncing solutions.  Although bouncing 
solutions were found in 
ref.~\cite{Copeland:vi,Easther:1995ba,Copeland:1997ug} by including the 
axion field in the tree-level action, this corresponds to the case where 
the string coupling and the scalar curvature diverge in both past and 
future asymptotics.  Therefore this can not be regarded as ideal 
singularity-free bouncing solutions unless some higher-order effects alter 
the dynamics of the system when the dilaton enters the strong coupling 
regime.

We can take into account the radiation-like matter source which comes,
for example, from the ideal string-gas \cite{Bassett:2003ck} or the 
5-form of the Ramond-Ramond sector of the type 
IIB superstring \cite{Constantinidis:1999cu,Fabris:2002jt,Fabris:2002pm}.  
In the presence of the radiation matter and the axionic field in $d=3$ 
spatial dimensions, Constantinidis {\it et al.} \cite{Constantinidis:1999cu} 
showed that bouncing solutions can be obtained when the dilatonic Brans-Dicke 
parameter $\omega$ is negative ({\it i.e.,} including the $\omega=-1$ case).  
When $-4/3<\omega<0$ these solutions exhibit divergent behavior of the scalar 
curvature as the dilaton evolves toward the large coupling regime ($\vp \to 
\infty$), in which case Constantinidis {\it et al.} did not regard them 
as ideal nonsingular bouncing solutions.  Meanwhile these solutions, in 
another asymptotic, tend to approach the radiation dominant universe with a 
decreasing curvature and a finite dilaton.  Then it should be possible to 
have regular bouncing solutions if the initial state of the universe is in 
the weakly coupled region with a finite curvature as in the case of 
string/brane-gas cosmology.

In this paper we shall investigate singularity-free cosmological scenarios 
based on the effective action of the dilaton-modulus-axion system in the 
presence of the matter source term.  We keep the dilatonic Brans-Dicke 
parameter $\omega$ arbitrary so that the effective action includes a wide 
variety of theories such as the 
$F$-theory \cite{Vafa:1996xn,Khviengia:1997rh} or the multidimensional 
theory \cite{Bronnikov:2002ki}.  Indeed it was recently found that in the 
presence of the axion and the pure radiation complete regular bouncing 
solutions were found in both past and future asymptotics when $\omega<-3/2$ 
in the case of $d=3$ spatial dimensions \cite{Fabris:2002jt,Fabris:2002pm}.  
With an application to string/brane-gas cosmology in mind, we shall 
extend the analysis to the general $d$ spatial dimensions with/without the 
axion and modulus fields.  In fact there exist a number of bouncing 
solutions even for the tree-level action in the presence of the radiation 
source term.  In particular we will make numerical analysis in the context 
of string-gas cosmology with and without the axion by including the 
massive state together with the leading massless state (radiation) and 
demonstrate the existence of nonsingular bouncing solutions even for 
$\omega=-1$.

When the system evolves toward the strongly coupled regime, it is likely 
that the evolution is altered by the higher-order corrections to the 
tree-level action.  Therefore we shall also make numerical simulations for 
the case where the dilatonic high-order derivative and loop corrections are 
included in the dilaton-modulus-axion system with a matter source term.  We 
will show that these higher-order corrections typically dominate the 
dynamics of the system, by which the big-bang singularity can be avoided as 
in the case where the only dilaton is present.

\section{Model}

Consider the following effective action with three scalar fields
$\vp$, $\chi$ and $\sigma$: 
\begin{eqnarray}
S=\int d^D x \sqrt{-g_D} \left[ \frac{e^{-\varphi}}{2} 
\left\{R-\omega (\nabla \varphi)^2-(\nabla \chi)^2 
-e^{-(n-1)\varphi} (\nabla \sigma)^2 
\right\}+ {\cal L}_m +{\cal L}_c \right]\,,
\label{Baction}
\end{eqnarray}
where $R$ is the scalar curvature in $D$-dimensions with metric 
$g_{\mu\nu}$ with $g_D={\rm det}\,g_{\mu\nu}$. 
${\cal L}_m$ represents the matter source term and ${\cal L}_c$ corresponds
to the higher-order corrections to the tree-level action.  The action 
(\ref{Baction}) contains a variety of models derived from string theories.  
In what follows we shall briefly show such examples.

\begin{itemize}
\item {\em Low energy effective string action}

The lowest order $\bar{D}$-dimensional string effective action 
coming from the beta function of the string world-sheet is characterized by 
\cite{Lidsey:1999mc} 
\begin{eqnarray}
S_{\bar{D}}=\int d^{\bar{D}} x \sqrt{-g_{\bar{D}}} \,
\frac{e^{-\psi}}{2} \left[ R_{\bar{D}}+(\nabla \psi)^2-\frac{1}{12} H_{\mu 
\nu \lambda} H^{\mu \nu \lambda} \right]\,,
\label{lowest}
\end{eqnarray}
where $\psi$ is the dilaton, $H_{\mu \nu \lambda}$ is the antisymmetric
tensor field.  Let us compactify the $k=\bar{D}-4$ dimensions by 
introducing a modulus field, $e^{k\bar{\chi}}$.  We define an effective 
dilaton in four dimensions as $\vp=\psi-k \bar{\chi}$. We also introduce a
pseudo-scalar axion field $\sigma$ as 
$H^{\mu\nu\lambda}=e^{\vp}\epsilon^{\mu\nu\lambda \rho} \nabla_{\rho} 
\sigma$ by taking into account the fact that $H^{\mu\nu\lambda}$ has only one 
degree of freedom in 4 dimensions \cite{Lidsey:1999mc}.  
Then we get the action (\ref{Baction}) 
with $D=4$, $\omega=-1$, $\chi=\sqrt{k}\bar{\chi}$, $n=-1$, ${\cal L}_m=0$ 
and ${\cal L}_c=0$.
 
\item {\em String/Brane-gas cosmology}

The string-gas cosmology \cite{Bassett:2003ck} is an attempt to explain why 
and how the three spatial dimensions become large and other 6 or 7 are kept 
to be small.  The bulk action in 10 or 11 dimensions is assumed to take the 
same form as (\ref{lowest}), but we have additional source matters due to 
the presence of the ideal string-gas in thermal equilibrium.  The simplest 
version of the string-gas cosmology \cite{Bassett:2003ck} corresponds to 
$\omega=-1$, $\chi=0$, $\sigma=0$, ${\cal L}_m \ne 0$ and ${\cal L}_c=0$ in 
(\ref{Baction}).  Instead of introducing a modulus field, one considers the 
evolution of one scale factor for $(D-1)$ spatial dimensions or two scale 
factors for large and small dimensions \cite{Bassett:2003ck}.  
In this work we shall also analyze 
the case where the axion is present ($\sigma \ne 0$).  Note that the 
brane-gas cosmology discussed in refs.~\cite{Alex,Brand} belongs to this 
class with a different matter source when we use the low energy bulk action 
(\ref{lowest}) with $\omega=-1$.

\item {\em Cosmology with $p$-brane or D$p$-brane solitons}

The $p$-brane or D$p$-brane are the solitonic degrees of freedom on which
string endpoints live.  These can be fundamentally important when the string 
coupling is large, since they are light in the high curvature regime 
\cite{Maggiore:1998cz,Riotto:1999kn}.  In the string $\sigma$-model whose 
metric is minimally coupled to the $p$-brane, Duff {\em et al.} 
\cite{Duff:1994an} showed that the effective action can be described by 
(\ref{Baction}) with $\chi=0$, $\sigma=0$, ${\cal L}_m \ne 0$ and 
\begin{eqnarray}
\omega=-\frac{(D-1)(\bar{p}-2)-\bar{p}^2}{(D-2)(\bar{p}-2)-\bar{p}^2}\,,
\label{omebrane}
\end{eqnarray}
where $\bar{p}=p+1$.  
{}From this we have $\omega=-1$ for the string ($p=1$).
However $\omega$ depends upon the values of $d$ and $p$
for the $p$-branes with $p \ne 1$. 
Cosmology in this scenario was investigated 
in refs.~\cite{Park:1997dw,Park:1999xn} in the absence of higher-order 
corrections to the tree-level action (see also ref.~\cite{Rama}).

\item {\em F-theory}

The superstring type IIB theory can be reformulated geometrically 
within the framework of 12 dimensional theory-- 
called $F$-theory \cite{Vafa:1996xn,Khviengia:1997rh}.  
After the dimensional reduction from 12 to 10 dimensions, the following 
action may be obtained \cite{Fabris:2002pm}:
\begin{eqnarray}
S=\frac12 \int d^{10} x \sqrt{-g_{10}}\left[ e^{-\vp}\left( R_{10}
+3(\nabla \vp)^2-\frac{1} {12} F_{\mu \nu \lambda} F^{\mu \nu \lambda} 
\right)-\frac{1}{8} F_{\mu\nu}F^{\mu\nu} \right]\,,
\label{F}
\end{eqnarray}
where the terms $F_{\mu\nu}$ and $F_{\mu\nu\lambda}$ come
from the Ramond-Ramond (RR) sector of the type IIB 
superstring.  This action is different from the standard string theory with 
$\omega=-1$.  Making the similar dimensional reduction from 10 to 4 
dimensions as (\ref{lowest}) with the isotropization of the RR terms, we 
get the action (\ref{Baction}) with $\omega =-3$, $\chi \ne 0$, $\sigma \ne 
0$ ($n=-1$) and the radiation fluid term (${\cal L}_m \ne 0$) coming from 
the 5-form of the RR sector.  The axionic terms minimally coupled to the 
dilaton can also appear from the same RR sector \cite{Fabris:2002pm}.  This 
corresponds to the case of adding another field with $n=0$ in the action 
(\ref{Baction}).

\item {\em Multidimensional cosmologies}

The low energy limit of the string effective action may be reformulated in the 
context of multidimensional theories \cite{Freund:1982pg}.  In its simplest 
form the multidimensional theories contain only the geometry, in which case 
one has $\omega=(1-\tilde{d})/\tilde{d}$ with $\chi=\sigma=0$ by the 
dimensional reduction ($\tilde{d}$ is the number of compactified 
dimensions).  When a 2-form field or a conformal gauge field with 
$(\tilde{d}+4)/2$-form is present, we have an axionic term $\sigma$ in 
eq.~(\ref{Baction}) with $n=-2/\tilde{d}+1$ or $n=-2/\tilde{d}$, 
respectively.  Therefore we keep the value of $n$ arbitrary so that 
multidimensional theories are also included in (\ref{Baction}).

\end{itemize}

Hereafter we shall analyze the cosmological solutions based
on the action (\ref{Baction}).  
For convenience we call the fields $\vp$, $\chi$ and $\sigma$
as dilaton, modulus and axion, respectively, unless otherwise specified.
We do not include the potential of scalar fields 
in the action\footnote{See ref.~\cite{Easson:1999xw} for the construction of 
nonsingular cosmological solutions including the dilaton potential.}.  
Extending the analysis of 
refs.~\cite{Constantinidis:1999cu,Fabris:2002jt,Fabris:2002pm} with $D=4$, 
we will study the cosmological evolution in the general $D$-dimensional 
action with a time-dependent modulus field ($\dot{\chi} \ne 0$).  In 
applying to string-gas or $p$-brane cosmologies, it is convenient to 
consider the case of the general dimension $D$.  In addition the presence 
of the internal space (modulus) can alter the parameter range where the 
nonsingular solutions exist.  We should also keep in mind that higher-order 
corrections to the tree-level action are inevitably important in the 
strongly coupled regime ($e^{\vp}~\gsim~1$).  In later sections we shall 
numerically investigate the dynamics of the system in the action 
(\ref{Baction}) when the dilatonic $\alpha'$ and loop corrections are taken 
into account (${\cal L}_c \ne 0$).

Let us consider the case where the $D=d+1$ dimensional spacetime
is described by a flat Friedmann-Robertson-Walker (FRW) metric 
with the line element 
\begin{eqnarray}
ds^2=-dt^2+a^2(t) \delta_{ij}dx^i dx^j\,,
\label{line}
\end{eqnarray}
where $a(t)$ is the scale factor.
In the context of the string (brane)-gas cosmology, it is possible to consider 
the case with two scale factors corresponding to the ``large'' and 
``small'' dimensions.  It was shown in ref.~\cite{Bassett:2003ck}
that only the large dimensions can be dynamically important whereas
the small dimensions are kept to be nearly constant.
In such cases it is sufficient to analyze the case of the one scale factor
described by eq.~(\ref{line}).  However we will also discuss the case of two 
scale factors in sec.~5.

Defining a new scalar field, 
\begin{eqnarray}
\phi=e^{-\vp}\,,
\label{phidef}
\end{eqnarray}
the variation of the action (\ref{Baction}) yields the following 
equations of motion 
\begin{eqnarray}
\label{b1}
 & &H^2 = \frac{1}{d(d-1)\phi} \left( \omega \frac{\dot\phi^2}{\phi}
+\phi \dot{\chi}^2+\phi^n \dot{\sigma}^2 -2dH\dot{\phi}
+2\rho+2\rho_c \right) \,, \\
\label{b2}
& &\dot{H} = -\frac{1}{(d-1)\phi}\left(\omega 
\frac{\dot\phi^2}{\phi}+\phi \dot{\chi}^2+\phi^n \dot{\sigma}^2
-H\dot{\phi}+\ddot{\phi}+\rho+p+\rho_c+p_c \right)\,, \\
\label{b3}
& & \ddot{\phi}+dH\dot{\phi}-\frac{\phi}{2\omega} 
\left( \omega \frac{\dot\phi^2}{\phi^2}+\dot{\chi}^2
+n\phi^{n-1}\dot{\sigma}^2+R+\Delta_c \right)=0\,, \\
\label{b4}
& & \ddot{\chi}+\left(dH+\frac{\dot{\phi}}{\phi}\right) \dot{\chi}=0\,, \\
\label{b5}
& & \ddot{\sigma}+\left(dH+n\frac{\dot{\phi}}{\phi}\right) 
\dot{\sigma}=0\,, \\
\label{b6}
& & \ddot{\rho}+dH(\rho+p)=0\,,
\label{back}
\end{eqnarray}
where $H \equiv \dot{a}/a$ is the Hubble rate. 
Here $\rho$ and $p$ stand for the energy and pressure density of the matter
source, respectively.  $\rho_c$ and $p_c$ correspond to the dilatonic 
higher-order corrections to the tree-level action with stress-energy tensor 
$T^{\mu}_{\nu}=(-\rho_c, p_c, p_c, p_c)$.  $\Delta_c$ comes from the 
variation of ${\cal L}_c$ with respect to $\phi$.

The scalar curvature, $R$, is expressed as 
\begin{eqnarray}
R &=& d(d+1)H^2+2d\dot{H} \\
&=& -\omega \frac{\dot\phi^2}{\phi^2}-\dot{\chi}^2-
\phi^{n-1} \dot{\sigma}^2+\frac{2}{(d-1)\phi}
\left(-d^2H\dot{\phi}-d\ddot{\phi}+\rho-dp
+\rho_c-dp_c \right)\,.
\label{scalar}
\end{eqnarray}
Substituting this relation for eq.~(\ref{b3}), we get
\begin{eqnarray}
\ddot{\phi}+dH\dot{\phi}+
\frac{(1-n)(d-1)\phi^n \dot{\sigma}^2-2(\rho-dp)
-2(\rho_c-dp_c) -(d-1)\phi \Delta_c}{2\{\omega(d-1)+d\}} 
=0\,.
\label{phieq0}
\end{eqnarray}
The equations (\ref{b4}) and (\ref{b5}) are easily integrated 
to give 
\begin{eqnarray}
\dot{\chi}=\frac{A}{a^d\phi}\,,~~~~ 
\dot{\sigma}=\frac{B}{a^d\phi^n}\,,
\label{ana}
\end{eqnarray}
where $A$ and $B$ are integration constants. 
Making use this relation and 
introducing a new time parameter, 
\begin{eqnarray}
\tau \equiv \int a^{-d} dt \,,
\label{tau}
\end{eqnarray}
the $\phi$ equation (\ref{phieq0}) can be written in the form: 
\begin{eqnarray}
\frac{{\rm d}^2 \phi}{{\rm d} \tau^2}+ 
\frac{(1-n)(d-1)B^2}{2\{\omega(d-1)+d\}}\phi^{-n} 
-\frac{2(\rho-dp)+2(\rho_c-dp_c)+(d-1)\phi 
\Delta_c}{2\{\omega(d-1)+d\}} a^{2d}=0 \,.
\label{phieq}
\end{eqnarray}
In the absence of the correction ${\cal L}_c$, the last term of 
eq.~(\ref{phieq}) vanishes for $\rho=p=0$ (no matter source) or $\rho=dp$ 
(radiation).  In these cases it is possible to obtain analytic solutions 
for the general spatial dimension $d$ even in the presence of axion and 
modulus fields.  Notice that the simplest pre-big-bang scenario 
\cite{Veneziano:1991ek} corresponds to the case of $\rho=p=0$ without 
the axion field.  The radiation matter ($\rho=dp$) can appear naturally 
in the context of string-gas cosmology at finite temperature 
\cite{Bassett:2003ck} or the 5-form existing in the RR sector of the type 
IIB superstring \cite{Fabris:2002pm}.  When $\rho=dp$, the integration of 
eq.~(\ref{b6}) gives $\rho=\rho_0 a^{-(d+1)}$, with $\rho_0$ being a 
constant.  Introducing a new parameter, $b=a\phi^{1/(d-1)}$, eq.~(\ref{b1}) 
can be written for ${\cal L}_c =0$ as 
\begin{eqnarray}
d(d-1)\phi^2 \left(\frac{b'}{b}\right)^2=2\rho_0 b^{d-1}+ 
A^2+B^2\phi^{1-n}+\frac{\omega(d-1)+d}{d-1}\phi'^2\,,
\label{b}
\end{eqnarray}
where a prime denotes the derivative with respect to $\tau$. 
This equation includes the case without the matter source by setting 
$\rho_0=0$.

In the next two sections we shall derive the analytic solutions for the scale 
factor and the dilaton in the case of the radiation or no matter source 
without the correction ${\cal L}_c$.  For the general equations of state 
with $p=\gamma \rho$, it is not so easy to find analytic solutions due to the 
fact that the last term in eq.~(\ref{phieq}) does not vanish\footnote{Note, 
however, that it is still possible to obtain analytic solutions for some 
special equations of state in the absence of the axion and the modulus 
\cite{Park:1997dw}.}.  In sec.~5 we will analyze the dynamics of 
string-gas cosmology as an application by including the massive state 
(Kaluza-Klein and winding modes) in addition to the leading massless state 
(radiation), with and without the axion.  In this case the string-gas state 
is not purely the radiation.  Since it is difficult to proceed analytic 
approach when the higher-order correction ${\cal L}_c$ is present, we shall 
make numerical analysis separately in sec.~6 for the general action 
(\ref{Baction}).  Hereafter we will consider the case of $d>1$.

\section{Low energy tree-level action without axion 
($\sigma=0$ and ${\cal L}_c=0$)} 

Let us first analyze the low energy tree-level action without 
the axion (${\cal L}_c=0$ and $\sigma=0$).  When $\rho=p=0$ (no matter 
source) or $\rho=dp$ (radiation), eq.~(\ref{phieq}) is integrated to give 
\begin{eqnarray}
\phi=c\tau\,,
\label{phino}
\end{eqnarray}
where $c$ is a constant. 
We can set another integration constant to be zero without 
loss of generality. Since $\phi=e^{-\vp}$ is required to be larger than zero, 
we have $\tau>0$ for $c>0$ and $\tau<0$ for $c<0$.  The string coupling, 
$g_s^2=e^{\vp}=\phi^{-1}$, diverges as $\tau \to 0$.

Since we are now considering the case with $B=0$,
eq.~(\ref{b}) yields
\begin{eqnarray}
\int \frac{\rd b}{b\sqrt{Mb^{d-1}+N}}=\pm 
\int \frac{\rd \tau}{\phi}\,,
\label{int}
\end{eqnarray}
where 
\begin{eqnarray}
M = \frac{2\rho_0}{d(d-1)}\,,~~~~ 
N = \frac{1}{d(d-1)}\left[A^2+ \frac{\omega (d-1)+d}{d-1}
c^2 \right]\,.
\label{MN}
\end{eqnarray}
The integral in the l.h.s. of eq.~(\ref{int}) is different depending on 
the sign of $N$.
For example, $N$ is positive in the standard string theory ($\omega=-1$) with 
$d>1$.  However, $N$ can be negative for the theories such as the solitonic 
$p$-brane/D$p$-brane or multidimensional theories.  In what follows we 
shall discuss the positive and negative $N$ cases separately.

\subsection{Case of $N>0$}

When $N>0$ the integration of eq.~(\ref{int}) leads to 
\begin{eqnarray}
\frac{b^{d-1}}{\left(\sqrt{Mb^{d-1}+N}+\sqrt{N}
\right)^2}=\frac{1}{\xi^2}\,.
\label{pN}
\end{eqnarray}
where
\begin{eqnarray}
\xi=\xi_0 \left(c\tau\right)^q\,~~~~ 
{\rm with}~~~~q=\pm 
\frac{\sqrt{N}}{2c}(d-1)\,.
\label{xiq}
\end{eqnarray}
Here $\xi_0$ is a positive integration constant.
Then the analytic solution for the scale factor can be written 
as 
\begin{eqnarray}
\label{ageneral}
a &=& \left(\frac{2\sqrt{N}}{\sqrt{\phi}(\xi-M/\xi)} \right)^{2/(d-1)} 
\\
\label{anoB}
&=& f(\phi)^{2/(d-1)}\,~~~{\rm with}~~~
f(\phi)=\frac{2\sqrt{N}}{\xi_0\phi^{\frac12+q}- 
(M/\xi_0)\phi^{\frac12-q}} \,.
\end{eqnarray}
This indicates that the evolution of the scale factor is determined
in terms of the field $\phi$ for the case (\ref{phino}). 
The simplest pre-big-bang scenario with a constant modulus 
corresponds to the case of $A=0$, $B=0$, $M=0$,
$d=3$ and $\omega=-1$, thereby 
yielding $q=\pm \sqrt{3}/6$ and 
\begin{eqnarray}
a=\frac{2\sqrt{N}}{\xi_0} \phi^{(-3\pm\sqrt{3})/6}
\,.
\label{PBB}
\end{eqnarray}
This represents an inflationary solution with decreasing $\phi$ from the 
weakly coupled regime ($\phi \to \infty$) to the strongly coupled regime 
($\phi \to 0$).  Note that the solution eventually reaches the curvature 
singularity as $\phi \to 0$.  If we consider the case of increasing $\phi$, 
the solution presented in eq.~(\ref{PBB}) describes a contracting universe.  
In this case the solution does not connect to our expanding branch.

The situation changes when the radiation is taken into account ($M \ne 0$).
Differentiating eq.~(\ref{anoB}) with respect to $\phi$, we have 
\begin{eqnarray}
\frac{{\rm d} f}{{\rm d} \phi} (\phi)=-\sqrt{N} \frac{(1+2q)\xi_0 
\phi^{\frac12+q}-(1-2q)(M/\xi_0)\phi^{\frac12-q}} {(\xi_0 
\phi^q-(M/\xi_0)\phi^{-q})^2} \,.
\label{diffe}
\end{eqnarray}
This means that it is possible to have bouncing solutions with the increase of 
$\phi$ as long as $-1/2<q<0$.  In this case, if the condition $(1+2q)\xi_0 
\phi^{\frac12+q}> (1-2q)(M/\xi_0)\phi^{\frac12-q}$ is satisfied initially 
with a finite value of the dilaton ($\phi_i \ne 0$), the universe contracts 
at the initial stage.  This is followed by a bounce due to the growth of 
the $(1-2q)(M/\xi_0)\phi^{\frac12-q}$ term.  The bounce occurs when the 
field $\phi$ passes at 
\begin{eqnarray}
\phi_*=\left(\frac{1-2q}{1+2q}\frac{M}{\xi_0^2}
\right)^{1/(2q)} \,.
\label{phistar}
\end{eqnarray}
The universe begins to expand for $\phi>\phi_*$ with the increase of $\phi$.  
As found from eq.~(\ref{anoB}) the scale factor goes to infinity as $\phi$ 
approaches the critical value, 
\begin{eqnarray}
\phi_c=\left(\frac{M}{\xi_0^2}\right)^{1/(2q)} \,.
\label{phicri}
\end{eqnarray}
Let us find the asymptotic behavior of $a$ around $\phi=\phi_c$.  Setting 
$\phi=\phi_c-c\Delta \tau$ with $\Delta \tau$ being a small time parameter, 
we find the scale factor (\ref{anoB}) is proportional to $(\Delta 
\tau)^{-2/(d-1)}$.  Since the cosmic time $t$ is related with $\tau$ by the 
relation $\rd t/\rd \tau=a^d$, the evolution of the scale factor is 
described by 
\begin{eqnarray}
a \propto t^{2/(d+1)}~~~({\rm for}~~~\phi \to \phi_c) \,,
\label{aasy}
\end{eqnarray}
which shows the radiation dominant behavior.

\begin{figure}
\epsfxsize = 4.5in \epsffile{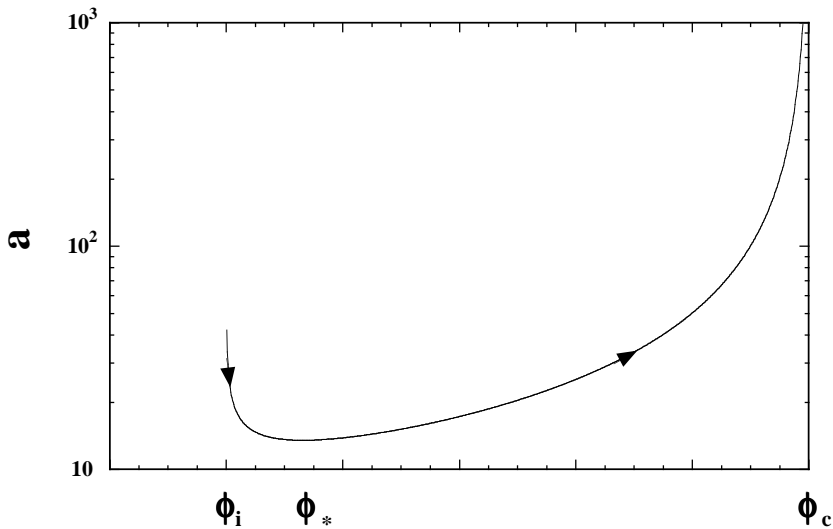} 
\caption{A nonsingular bouncing solution in the low energy tree-level action 
without the axion in the presence of the radiation ($\rho=dp$).  With the 
growth of $\phi=e^{-\vp}$, the universe contracts and exhibits the bounce at 
$\phi=\phi_*$.  After that the scale factor begins to grow and the system 
asymptotically approaches the radiation dominant universe as $\phi \to 
\phi_c$.}
\label{mlessevo}
\end{figure}

In the limit of $\phi \to \phi_c$, we have $t \to \infty$ as $\Delta
\tau \to 0$.  Then the scalar curvature asymptotically 
approaches zero as $R \propto t^{-2}$ for $\phi \to \phi_c$.  In addition the 
string coupling, $g_s^2=1/\phi$, is also finite in this case.  
This means that nonsingular bouncing solutions with finite curvature 
and dilaton can be obtained by taking into account 
the radiation in the low energy tree-level string action.  
Notice that we are not considering the regime around the strong coupling 
limit ($\phi \to 0$).  The universe is assumed to start out from 
the region with finite dilaton and curvature, as in the string/brane-gas 
cosmologies.  Then the dilaton evolves toward the weakly coupled region
(see fig.~\ref{mlessevo}), which asymptotically approaches the radiation 
dominant universe given by (\ref{aasy}).  
In sec.~5 we will numerically investigate the existence of such 
bouncing solutions in string-gas cosmology taking into account the 
massive state in addition to the leading massless state.

These bouncing solutions exist for $-1/2<q<0$, whose condition is written as 
\begin{eqnarray}
\omega<-\frac{A^2}{c^2} \,.
\label{boucon}
\end{eqnarray}
In the absence of the modulus, this condition yields $\omega<0$.  
Note, however, that the large initial value of the modulus 
restricts the range for the existence of the bouncing solutions.

If the condition, $(1+2q)\xi_0 
\phi^{\frac12+q}<(1-2q)(M/\xi_0)\phi^{\frac12-q}$, is satisfied initially 
with $-1/2<q<0$, the scale factor (\ref{anoB}) grows from the 
beginning with the increase of $\phi$.  Then it asymptotically approaches 
the radiation dominant solution given by eq.~(\ref{aasy}). 
This solution was numerically found as well in the context of string-gas 
cosmology \cite{Bassett:2003ck}.  Note that when $|q| \ge 1/2$ we have 
either ever expanding or contracting solutions.

It is possible to obtain bouncing solutions provided the condition 
(\ref{boucon}) is satisfied and the dilaton $\vp$ decreases.
In contrast, it is worth mentioning the case where the dilaton evolves toward 
the strongly coupled regime.  In this case, the scalar curvature 
does not necessarily converge in the limit of $\phi \to 0$.  Around the 
region of $\phi=0$, the scale factor is estimated as $a \propto 
\phi^{-\frac{1+2q}{d-1}}$ for $-1/2<q<0$.  Combining this relation with 
$\rd t/\rd \tau=a^d$, we have 
\begin{eqnarray}
t=t_0+\tilde{c}\tau^{-\frac{1+2dq}{d-1}} \,,
\label{texplicit2}
\end{eqnarray}
where $t_0$ and $\tilde{c}$ are constants.
Then the scale factor around $\phi=0$ is expressed as 
\begin{eqnarray}
a \propto (t-t_0)^{\frac{1+2q}{1+2dq}}\,.
\label{scaleasy2}
\end{eqnarray}
When $q>-1/(2d)$, one has $t \to \infty$ as $\tau \to 0$ from
eq.~(\ref{texplicit2}). Since the power in eq.~(\ref{scaleasy2}) is positive 
in this case, the universe expands for $\tau \to 0$.  
Meanwhile we have $t \to t_0$ as 
$\tau \to 0$ for $q<-1/(2d)$.  Since we are considering the case of 
$q>-1/2$, the power in eq.~(\ref{scaleasy2}) is negative for 
$-1/2<q<-1/(2d)$, thereby leading to the growth of $a$ as $t \to t_0$.

Since the scalar curvature evolves as $R \propto (t-t_0)^{-2}$ for 
the solution (\ref{scaleasy2}), it asymptotically decreases toward zero only 
for $q>-1/(2d)$.  When $-1/2<q<-1/(2d)$, $R$ diverges in the limit of  
$\phi \to 0$.  Therefore, 
when the system approaches the strongly coupled regime, we have bouncing 
solutions with finite curvature for $q>-1/(2d)$, {\em i.e.,} 
\begin{eqnarray}
\omega<-1-\frac{1}{d}-\frac{A^2}{c^2}\,.
\label{omef}
\end{eqnarray}
This does not include the string case ($\omega=-1$).
Combining (\ref{omef}) with the condition $N>0$,
we have $-\frac{d}{d-1}-\frac{A^2}{c^2}<\omega<
-1-\frac{1}{d}-\frac{A^2}{c^2}$.  When $d=3$ and $A=0$ (no modulus),
this condition yields $-3/2<\omega<-4/3$.

 We have to keep in mind that the description of the tree-level action 
 ceases to be valid as $\phi$ evolves toward the strong coupling region 
 ($\phi \to 0$).  The higher-order corrections to the tree-level action are 
 expected to be important in this regime.  On the contrary, nonsingular 
 bouncing solutions toward the weakly coupled regime shown in 
 fig.~\ref{mlessevo} can be regarded as ideal ones.  In addition the dilaton 
 and the scalar curvature do not diverge, provided that the universe starts 
 out from the region with finite $\vp$ and $R$.

\subsection{Case of $N<0$}

Let us next consider the case of $N<0$.  For example, when $d=3$ and $A=0$, 
this condition corresponds to $\omega<-3/2$.  The only difference relative 
to the $N>0$ case is the integral in the l.h.s.  of eq.~(\ref{int}).  For 
negative $N$ we have 
\begin{eqnarray}
{\rm arc tan}\sqrt{\frac{Mb^{d-1}+N}{|N|}}=
{\rm log}\, \xi_0 |c\tau+\phi_0|^q~~~
{\rm with}~~~q=\pm \frac{\sqrt{|N|}}{2c}(d-1)\,,
\label{ne}
\end{eqnarray}
where $\xi_0$ is a positive constant.
Then the scale factor can be written as 
\begin{eqnarray}
a=\left(\sqrt{\frac{|N|}{M}} 
\frac{1}{\sqrt{\phi}\cos \left({\rm log}\,\xi_0 \phi^q 
\right)}\right)^{2/(d-1)} \,.
\label{scale2}
\end{eqnarray}
This indicates that nonsingular bouncing solutions exist in the interval
\begin{eqnarray}
\frac{1}{\xi_0} \exp \left[\left(2j-\frac12 \right) \pi \right]<\phi^q< 
\frac{1}{\xi_0} \exp \left[\left(2j+\frac12 \right) \pi \right]\,,
\label{interval}
\end{eqnarray}
where $j$ is an integer.
From eq.~(\ref{scale2}) we find that the bounce occurs at 
$\cos ({\rm log}\,\xi_0\phi_*^q)=q+\sqrt{q^2+1}$, {\em i.e.,} 
\begin{eqnarray}
\phi_*=\frac{1}{\xi_0^{1/q}} \exp \left[\frac{1}{q}
{\rm arc} \cos \left(q+\sqrt{q^2+1}\right) \right]\,.
\label{phistar2}
\end{eqnarray}
Then we have $q<0$ from the condition, $|\cos ({\rm log}\,\xi_0\phi_*^q)| 
\le 1$. The asymptotic behavior of the scale factor is 
given by $a \propto t^{2/(d+1)}$ as the dilaton approaches $\phi \to 
(1/\xi_0^{1/q})\exp[(2j\pm 1/2)\pi/q]$.  In addition it is easy to show 
that the scalar curvature asymptotically decreases in the limit of $\phi 
\to (1/\xi_0^{1/q})\exp[(2j\pm 1/2)\pi/q]$.  Therefore the bouncing solution 
(\ref{scale2}) is characterized by the radiation dominant universe in both 
asymptotics with finite values of the dilaton and the scalar curvature.  We 
have a sequence of solutions (\ref{interval}) corresponding to the each 
integer value of $j$.  This is different from the case of $N>0$ where 
singularity-free solutions with finite dilaton exist only for 
$\phi_i<\phi<\phi_c$ ($\phi_i$ is the initial value).  The key point is 
that the dilaton can be controlled without entering the strongly coupled 
region with $\phi \sim 0$.

In the absence of the axion with $\rho=dp$ (or $\rho=p=0)$ considered above, 
the dilaton evolves as the free massless field ($\phi''=0$).  
However the situation changes when the axion is not dynamically negligible.

\section{Low energy tree-level action with axion 
($\sigma \ne 0$ and ${\cal L}_c=0$)} 

In this section we implement the effect of the axion and discuss
the difference of the dynamics compared to the case of $\sigma=0$.  
First, the evolution of the dilaton is altered by the existence of the second 
term in eq.~(\ref{phieq}).  When ${\cal L}_c=0$ and $\rho=dp$ (or 
$\rho=p=0$), we find a conserved quantity, $E$, from eq.~(\ref{phieq}) : 
\begin{eqnarray}
\frac12 \phi'^2+\frac{(d-1)B^2}{2\{\omega(d-1)+d\}} 
\phi^{1-n}=\frac{E}{2} \,.
\label{E}
\end{eqnarray}
Making use of this relation, eq.~(\ref{b1}) is reduced to the integral form 
(\ref{int}) with 
\begin{eqnarray}
M = \frac{2\rho_0}{d(d-1)}\,,~~~~ 
N = \frac{1}{d(d-1)}\left[A^2+ \frac{\omega (d-1)+d}{d-1}
E \right]\,.
\label{MN2}
\end{eqnarray}
The only difference relative to the $\sigma=0$ case is that $c^2$ in 
eq.~(\ref{MN}) is replaced for $E$.

The integral in the l.h.s. of eq.~(\ref{int}) can be performed as in the 
previous section depending on the sign of $N$.  In order to integrate the 
r.h.s. of eq.~(\ref{int}), it is required to make further classifications 
depending on the sign of the second term in eq.~(\ref{E}).

\subsection{$\omega>-d/(d-1)$}

When $\omega>-d/(d-1)$, it is convenient to introduce a new time parameter, 
$\eta$, satisfying 
\begin{eqnarray}
\label{phis}
\phi=\phi_0 \left(\sin \eta \right)^{2/(1-n)}\,~~~~{\rm with}
~~~~\phi_0=
\left(\frac{E\{\omega (d-1)+d\}}{(d-1)B^2}\right)^{1/(1-n)}\,.
\label{nepx}
\end{eqnarray}
{}From eq.~(\ref{E}) we find the relation 
$\rd \phi/\rd \tau= \pm \sqrt{E}\cos \eta$.  
Together with eq.~(\ref{MN2}) we obtain 
\begin{eqnarray}
\frac{\rd \tau}{\rd \eta}=\pm \frac{2}{(1-n)\sqrt{E}}\phi_0
(\sin \eta)^{\frac{1+n}{1-n}}\,.
\label{taueta}
\end{eqnarray}
Making use of this relation it is easy to perform the integral 
(\ref{int}).  
Notice that we are now considering the case with $N>0$
since this is automatically satisfied for  $\omega>-d/(d-1)$.
The scale factor is expressed by the form (\ref{ageneral}) with 
\begin{eqnarray}
\xi&=&\xi_0 \left| {\rm tan}\frac{\eta}{2}\right|^q\,,~~~~
q=\pm \frac{d-1}{1-n}\sqrt{\frac{N}{E}}\,.
\label{xp}
\end{eqnarray}
Namely the explicit form of $a$ can be written in terms of $\eta$ :
\begin{eqnarray}
a=\left(\frac{2\sqrt{N}}{\sqrt{\phi_0}(\sin \eta)^{1/(1-n)}
\left\{\xi_0 |\tan \eta/2|^q-(M/\xi_0)|\tan \eta/2|^{-q}\right\}} 
\right)^{2/(d-1)}\,.
\label{sca3}
\end{eqnarray}
The evolution of the scale factor is well understood by considering 
the behavior of $\xi$ and $\phi$ for $0<\eta<\pi$
(or the region $2j\pi<\eta<(2j+1)\pi$ with $j$ being an integer).

Let us first investigate the case of $n<1$ and $q<0$.  
It is sufficient to analyze negative values of $q$ only, since the function 
$\xi$ is symmetric with respect to $\eta=\pi/2$ for the change of $q \to 
-q$.  As found from eq.~(\ref{ageneral}), $\xi$ is larger than $\sqrt{M}$.  
It is clear from eq.~(\ref{ageneral}) that the universe expands as $\xi$ 
approaches $\xi_c \equiv \sqrt{M}$ (note that $\phi$ is finite in this 
limit).  Consider the asymptotic behavior of the scale factor when $\xi$ 
evolves toward $\xi_c=\xi_0 \left| {\rm tan} (\eta_c/2) \right|^q$ with $M 
\ne 0$.  Setting $\eta=\eta_c-\Delta \eta$ in eq.~(\ref{ageneral}), we find 
that the scale factor evolves as $a \propto (\Delta \eta)^{-\frac{2}{d-1}}$ 
around $\eta=\eta_c$.  Making use of the relation (\ref{sca3}) and $\rd 
\tau/\rd t=a^{-d}$, we have $a \propto t^{2/(d+1)}$ with $t \to \infty$ as 
$\Delta \eta \to 0$.  This represents the radiation dominant universe with 
decreasing curvature ($R \to 0$) as is similar to the case (\ref{aasy}).

In another limit $\eta \to 0$, one has $\phi \to 0$ and $\xi \to 
\infty$ for $n<1$ and $q<0$.  Since the evolution of the scale factor is 
approximately described by $a \propto 
\eta^{-\frac{2}{d-1}\left(\frac{1}{1-n}+q\right)}$ around $\eta=0$, we get 
an expanding solution for $q>-\frac{1}{1-n}$.  
{}From eq.~(\ref{MN2}) and (\ref{xp}) this condition yields 
\begin{eqnarray}
\omega<-\frac{A^2}{E} \,,
\label{ome1}
\end{eqnarray}
which is similar to eq.~(\ref{boucon}). 
Following the similar process by which eqs.~(\ref{texplicit2}) and 
(\ref{scaleasy2}) are derived, we get the evolution of the scale factor 
around $\eta=0$ as 
\begin{eqnarray}
a \propto (t-t_0)^{\frac{1+q(1-n)}{1+dq(1-n)}}\,.
\label{scaleasy}
\end{eqnarray}
Here $t=t_0+\tilde{c}\eta^{-\frac{2}{d-1}\left(\frac{1}{1-n}+dq \right)}$
with $t_0$ and $\tilde{c}$ being integration constants.
It is easy to show that the scalar curvature decreases toward zero
for $q>-\frac{1}{d(1-n)}$, whereas $R$ diverges for
$q<-\frac{1}{d(1-n)}$.
If we want to construct bouncing solutions where the dilaton evolves 
toward $\phi=0$ with finite curvature, this requires the condition 
$q>-\frac{1}{d(1-n)}$, {\em i.e.}, 
\begin{eqnarray}
\omega<-1-\frac{1}{d}-\frac{A^2}{E}\,,
\label{omecon}
\end{eqnarray}
which is analogous to eq.~(\ref{omef}).  
Combining this with the condition $\omega>-d/(d-1)$, we have 
$-3/2<\omega<-4/3$ for $d=3$ and $A=0$ (no modulus), as shown in 
ref.~\cite{Constantinidis:1999cu}.  Notice that the presence of the modulus 
makes the allowed range of $\omega$ narrower.

Again we should caution that the string coupling, $g_s^2=\phi^{-1}$, 
diverges in the asymptotic limit $\phi \to 0$.  
In this sense these solutions can not be 
regarded as ideals ones, unless some higher-order effects are taken into 
account in the strong coupling regime.  Meanwhile when the universe begins 
from a state with finite dilaton and curvature satisfying $\phi_i~\gsim~1$, 
it is possible to have nonsingular bouncing solutions provided that the 
solutions approach the region around $\xi=\xi_c$.  In this case the dilaton is 
finite for the range $\phi_i<\phi<\phi_c$ with $\phi_c=\phi_0 \left(\sin 
\eta_c \right)^{2/(1-n)}$.  The system asymptotically approaches the 
radiation dominant universe with decreasing curvature.  One different point 
relative to the $\sigma=0$ case is that the axion tends to work for the dilaton 
to turn back toward the strong coupling regime.  Nevertheless the string 
coupling does not diverge as long as the dilaton approaches the value 
$\phi=\phi_c$ instead of $\phi=0$.

When $n>1$, both $\xi$ and $\eta$ diverge either in the limit of $\eta \to \pi$
or $\eta \to 0$, depending on the sign of $q$.  Since the asymptotic 
evolution of the scale factor exhibits the contracting behavior, it is not 
possible to have bouncing solutions for $n>1$.

At the final of this subsection, it is worth mentioning the case without 
the matter source ($M=0$).  When $n<1$ and $q<0$, the scale factor approaches 
the asymptotic solution given by eq.~(\ref{scaleasy}) in both limits of 
$\eta \to 0$ and $\eta \to \pi$.  Therefore if the condition (\ref{omecon}) 
is not satisfied, the scalar curvature diverges in both past and future 
asymptotics together with the infinite string coupling.  The $\omega=-1$ 
case is contained in this class.  The low energy tree-level string action 
($\omega=-1$) with an axion leads to bouncing solutions as found in 
refs.~\cite{Copeland:vi,Easther:1995ba,Copeland:1997ug}, 
but these correspond to the solutions with divergent behavior of the dilaton 
and the curvature 
in past/future asymptotics.  When the condition (\ref{omecon}) is 
satisfied, the curvature asymptotically decreases but the string coupling 
diverges as $\eta \to 0$ and $\eta \to \pi$.  {}From these arguments, we 
find that inclusion of the matter source ($M \ne 0$) is crucial for the 
construction of nonsingular bouncing solutions with finite dilaton and 
curvature even in the presence of the axion.  In sec.~5 we will analyze the 
dynamics of the system with and without axion in the context of 
string-gas cosmology.

\subsection{$\omega<-d/(d-1)$}

Let us next proceed to the case of $\omega<-d/(d-1)$.
Introducing a new time parameter, $\eta$, defined by 
\begin{eqnarray}
\phi=\phi_0 (\sin {\rm h}\eta)^{2/(1-n)}\,~~~~
{\rm with}~~~~\phi_0=
\left(\frac{-E\{\omega (d-1)+d\}}
{(d-1)B^2}\right)^{1/(1-n)}\,,
\label{peta}
\end{eqnarray}
the relation between $\tau$ and $\eta$ yields 
\begin{eqnarray}
\frac{\rd \tau}{\rd \eta}=\pm \frac{2}{(1-n)\sqrt{E}}\phi_0
(\sin {\rm h}\eta)^{\frac{1+n}{1-n}}\,.
\label{taueta3}
\end{eqnarray}
In this case the value of $N$ in eq.~(\ref{MN2}) can be either
positive or negative\footnote{In the absence of the modulus, $N$
is negative.}.

\subsubsection{Case of $N>0$} 

The positive value of $N$ can be realized in the presence of the modulus. 
Integrating eq.~(\ref{int}), the scale factor is given by the 
form (\ref{ageneral}) with 
\begin{eqnarray}
\xi&=&\xi_0 \left| {\rm tan\,h}\frac{\eta}{2}\right|^q\,,~~~~ 
q=\pm \frac{d-1}{1-n}\sqrt{\frac{N}{E}}\,.
\label{xpd}
\end{eqnarray}
Consider the evolution of $a$ for $q>0$ with $\eta>0$.  
When $n<1$ both $\xi$ and $\phi$ increase as $\eta \to \infty$, 
in which case the universe contracts asymptotically.  
In the case of $n>1$, $\phi$ decreases toward zero, 
whereas $\xi$ approaches $\xi \to \xi_0$ as $\eta \to \infty$.  
In this asymptotic region, the scale factor evolves as 
\begin{eqnarray}
a \propto e^{\frac{2}{(n-1)(d-1)}\eta} 
\propto (t-t_0)^{\frac{2}{d+1-n(d-1)}}~~~~
{\rm with}~~~~t=t_0+\tilde{c} 
e^{\frac{(1+n)(d-1)-2d}{(1-n)(d-1)}\eta}\,,
\label{scal3}
\end{eqnarray}
where $t_0$ and $\tilde{c}$ are constants.  Since the power of the scale 
factor is larger than unity for $n>1$, this represents an inflationary 
solution.  Notice that this is achieved due to the presence of the modulus 
field. The curvature asymptotically decreases for $1<n<\frac{d+1}{d-1}$, 
whereas $R$ diverges for $n>\frac{d+1}{d-1}$.  
In both cases the string coupling 
diverges in the limit of $\eta \to \infty$.  The scale factor grows as the 
radiation dominant universe in another asymptotic, $\xi \to \sqrt{M}$.  In 
this limit the scalar curvature decreases toward zero with a finite string 
coupling.  Therefore it is possible to obtain nonsingular bouncing 
solutions for $n>1$ if the universe starts out from the region with finite 
dilaton and curvature and evolves toward the weakly coupled regime.  When 
the universe begins around the region $\xi \sim \sqrt{M}$ and evolves 
toward $\phi=0$, we also have the bouncing solutions that asymptotically 
approach the inflationary solution (\ref{scal3}).  However, since the 
solutions enter the strong coupling region in the latter case, this 
indicates the limitation of using the tree-level action.

When $q<0$ the scale factor approaches zero as $\eta \to \infty$
or $\eta \to 0$, depending on the sign of $q$.  This suggests a difficulty to 
obtain bouncing solutions for $q<0$.  In summary we can have 
regular bouncing solutions for $q>0$ and $n>1$.

\subsubsection{Case of $N<0$} 

When $N<0$ the integration of eq.~(\ref{int}) leads to
\begin{eqnarray}
a=\left(\sqrt{\frac{|N|}{M}} \frac{1}
{\sqrt{\phi} \cos \xi}\right)^{2/(d-1)}\,,
\label{scann}
\end{eqnarray}
where
\begin{eqnarray}
\xi&=& {\log }\left(\xi_0 \left| {\rm tan\,h}\frac{\eta}{2}\right|^q 
\right)\,,~~~~ q=\pm \frac{d-1}{1-n}\sqrt{\frac{|N|}{E}}\,.
\label{xpd2}
\end{eqnarray}
{}From this we find that bouncing solutions emerge in the interval 
$\eta_i<\eta<\eta_f$ with 
\begin{eqnarray}
\eta_i={\rm log}\left|
\frac{\xi_0^{1/p}+\exp\left\{(2j-1/2)\pi/p \right\}} 
{\xi_0^{1/p}-\exp\left\{(2j-1/2)\pi/p \right\}}\right|\,, ~~~~ 
\eta_f={\rm log}\left| \frac{\xi_0^{1/p}+\exp\left\{(2j+1/2)\pi/p \right\}} 
{\xi_0^{1/p}-\exp\left\{(2j+1/2)\pi/p \right\}}\right|\,.
\label{range}
\end{eqnarray}
Setting $\eta=\eta_i +\Delta \eta$ and $\eta=\eta_f -\Delta \eta$ around
$\eta=\eta_i, \eta_f$, the scale factor evolves as 
\begin{eqnarray}
a \propto (\Delta \eta)^{-2/(d-1)} \propto 
(t-t_0)^{\frac{n-1}{dn-1}} \,,
\label{scalasy}
\end{eqnarray}
where $t_0$ is a constant. 
When $n>1$ or $n<1/d$, the solution (\ref{scalasy}) represents the 
decelerating expansion of the universe with decreasing curvature ($R \to 
0$).  The standard axion coupling in $d=3$ dimensions ($n=-1$) is included 
in this case, and the solutions exhibit the radiation dominant behavior 
with $a \propto (t-t_0)^{1/2}$.  For the general axion coupling with 
$n \ne -1$, however, the evolution of the scale factor is not asymptotically 
radiation dominant.  This means that the axion affects the dynamics of the 
system even in the asymptotic limit.
In the case of $1/d<n<1$, we have bouncing solutions with growing curvature
in both asymptotics, which means that the solutions are not ideal ones.

We have many sequences of intervals with $\eta_i<\eta<\eta_f$, 
during each of which the string coupling and the curvature are kept finite. 
As in the case of $N<0$ with no axion, 
nonsingular bouncing solutions can be constructed without entering the strong 
coupling region around $\phi=0$.

\section{Application to string-gas cosmology}

In this section we shall consider the matter source term 
derived from the free energy of the ideal string-gas associated with type 
IIA/B string theory compactified on a square $T^9$-torus, simple product 
of nine circles.  It was recently shown that it is possible to construct 
the string-gas cosmological model where some dimensions ($d$) start to expand 
while the remaining dimensions are kept small \cite{Bassett:2003ck}.
In this work we include the effect of the axion.  We also wish to investigate 
the validity of the analytic solutions in the presence of the large and small 
dimensions.  In the region where the temperature, $T=1/\beta$, is below the 
Hagedorn one, $T_H=2\sqrt{2}\pi$, the free energy of a string-gas in the 
canonical ensemble takes the following form \cite{AO_Tan,Bassett:2003ck}: 
\bea F^{(d)}(\beta)=
-\frac{V_d}{2\pi\sqrt{\al}}\int_{-{1\over 2}}^{1\over
2}d\tau_1\int_0^{\infty}{d\tau_2\over\tau_2^{(3+d)/2}}\,
[\Lambda(r; \tilde{\tau})]^{9-d} \sum_{\tilde{p}=1}^\infty 
e^{-{\beta^2 \tilde{p}^2\over 4\pi\tau_2}}\, |M_2|^2(\tilde{\tau})\,,
\label{free}
\eea
where $V_d=a^d$ is the volume of the large $d$ dimensions with the scale 
factor $a$ and 
\be
\Lambda(r; \tilde{\tau})= \sum_{m_i,n_i=-\infty}^{+\infty} q^{\frac{1}{4}( 
{m_i\over r} + n_ir)^2}{\bar q}^{\frac{\al}{4}( {m_i\over r} - n_ir )^2}\,,
\label{Rlattice}
\ee
represents the contributions of the whole Kaluza-Klein (KK) and winding modes 
along the small dimensions.  Here $\tilde{\tau}$ is the modular parameter of 
the world-sheet torus and $q=\exp(2i\pi \tilde{\tau})$.  The sum in terms 
of $\tilde{p}$ in eq.~(\ref{free}) runs only over positive odd numbers and 
corresponds to taking the correct quantum statistic for bosons and 
fermions.

The $M_2$ factor in eq.~(\ref{free}) is expanded
in powers of $q$, as 
\beq
M_2(\tilde{\tau}) = 
\frac{\theta_2(\tilde{\tau})^4}{\tilde{\eta}(\tilde{\tau})^{12}} = 
\sum_{N_m=0}^{\infty} D(N_m)\,q^{N_m}\,,
\label{M2t}
\eeq
where $\theta_2$ and $\tilde{\eta}$ are modular functions 
on the torus (see ref.~\cite{PolBook}).  Here the value of $N_m$ corresponds 
to each string mass level with the degeneracy factor, $D(N_m)$. 
When we integrate the $\tau_1$ and $\tau_2$ integrals in eq.~(\ref{free}), 
it is convenient to extract the term with $N_m=\bar N_m=m_i=n_i=0$ that 
corresponds to the purely massless state.  Denoting this massless state 
(radiation) as $F_{rad}^{(d)}$, one has \cite{Bassett:2003ck} 
\be
F_{rad}^{(d)} = - \frac{a^d}{2\pi} D(0)^2 \Gamma
\left(\frac{d+1}{2}\right)(4\pi)^{(d+1)/2} \zeta(d+1) (1-2^{-(d+1)})
\beta^{-d-1}\,,
\label{freed0}
\ee
where $\Gamma$ and $\xi$ are the Gamma function and the Riemann 
zeta-function, respectively. 
{}From this the energy $E_{rad}^{(d)}$ and pressure 
$P_{rad}^{(d)}$ along the $d$ spatial dimensions are evaluated as 
\begin{eqnarray}
\label{E0rad}
E_{rad}^{(d)} &=& F_{rad}^{(d)}+\beta 
\frac{\partial F_{rad}^{(d)}}{\partial \beta} =
\frac{d a^d}{2\pi}D(0)^2 \Gamma \left(\frac{d+1}{2}\right) 
(4\pi)^{(d+1)/2} \zeta(d+1) (1-2^{-(d+1)}) \beta^{-d-1}\,, \\
\label{P0rad}
P_{rad}^{(d)} &=& -\frac1d\frac{\partial
F_{rad}^{(d)}}{\partial ({\rm ln}\, a)}= 
\frac{E_{rad}^{(d)}}{d} \,,
\end{eqnarray} 
which represents the equation of state for radiation in $d$ spatial
dimensions.  

The pressure along the  $(9-d)$-small dimensions is vanishing 
($P_{rad}^{(9-d)}=0$) for the pure massless state.  
In this case the evolution of the small 
dimensions is trivial and can be kept small around the duality symmetric 
radius, $r=1$ \cite{Bassett:2003ck}.  Therefore it is sufficient to 
consider the evolution of only the large $d$-dimensions with a constant 
radius of the small dimension for the pure massless state.  
Introducing the shifted dilaton, $\psi \equiv 
\varphi-d\mu$, with $a=e^{\mu}$, $\omega=-1$ and $\dot{\chi}=0$, 
eqs.~(\ref{b1})-(\ref{b3}) yield 
\begin{eqnarray}
\label{dilatoneq1}
-d\dot{\mu}^2+\dot{\psi}^2-e^{-(n-1)\vp}\dot{\sigma^2}
&=&e^{\psi}E\,, \\
\label{dilatoneq2}
\ddot{\mu}-\dot{\psi}\dot{\mu}-\frac{1-n}{2}e^{-(n-1)\vp}
\dot{\sigma}^2&=&\frac12 e^{\psi}P\,, \\
\label{dilatoneq3}
\ddot{\psi}-d\dot{\mu}^2-\frac{1+n}{2}e^{-(n-1)\vp}\dot{\sigma}^2
&=& \frac12  e^{\psi}E\,.
\label{reduce}
\end{eqnarray} 
Note that the energy and pressure are 
related with $\rho$ and $p$ in eqs.~(\ref{b1})-(\ref{b3}), as 
$E=2a^{d}\rho$ and $P=2a^{d}p$.
Since the pure massless state satisfies the equation of state with $\rho=dp$
from eq.~(\ref{P0rad}), the analytic method given in the previous section 
can be applied for this case.

\begin{figure}
\epsfxsize = 4.5in \epsffile{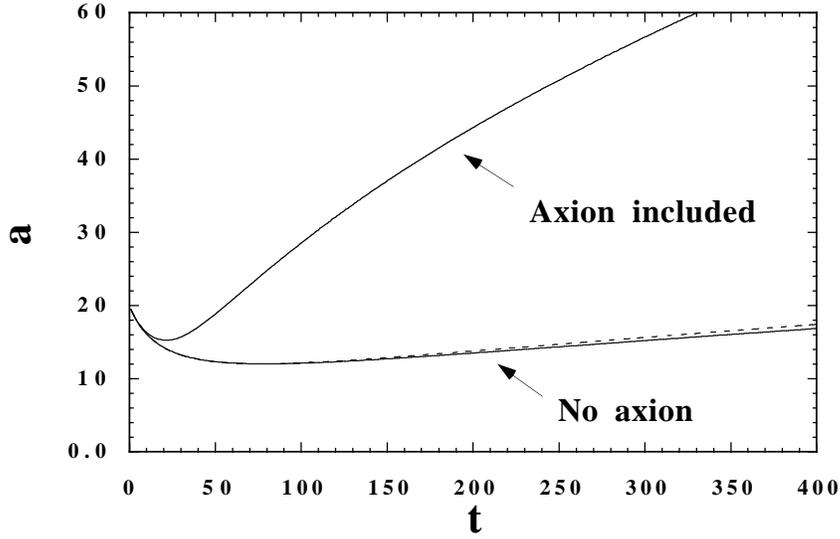}
\caption{The evolution of the scale factor $a$ in the case of $d=3$
with or without the axion for the initial conditions $\dot{\mu}_0=-0.03$, 
$\mu_0=3.0$, $\dot{\nu}_0=-0.005$, $\nu_0=0.01$, $\psi_0=-14$, 
$\beta_0=17$.  The massive effect is taken into account together with 
the radiation for the solid curves, whereas the dotted curve corresponds to 
the pure radiation case without the axion.  }
\label{scalef}
\end{figure}

\begin{figure}
\epsfxsize = 4.5in \epsffile{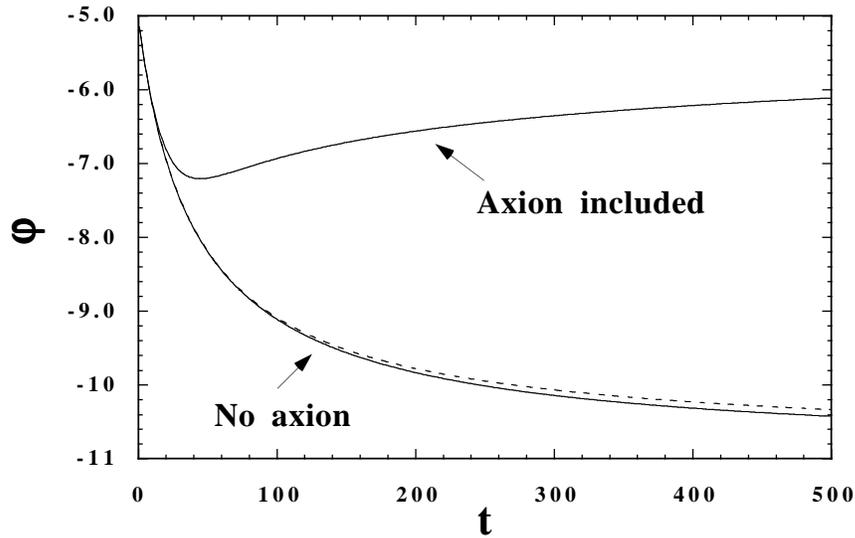}
\caption{The evolution of the dilaton $\varphi$ in the case of $d=3$ 
for the same initial conditions shown in fig.~\ref{scalef}. 
The massive effect is taken into account
together with the radiation for the solid curves, whereas the dotted curve 
corresponds to the pure radiation case without the axion.  }
\label{varphi}
\end{figure}

We are now considering the case of $\omega=-1$ and $A=0$, thereby
yielding a 
positive value of $N$ from eq.~(\ref{MN}).
In the absence of the axion we have $q=\pm \sqrt{3}/6$ from eq.~(\ref{xiq}).  
Since the negative value of $q$ satisfies $-1/2<q<0$, bouncing solutions 
can be obtained in this case as shown in sec.~3.1 [see fig.~\ref{mlessevo}].  
This is achieved by choosing negative finite initial values of $\dot{\mu}$, 
in which case the scalar curvature is also finite.  The scale factor 
exhibits a bounce at $\phi=\phi_*$ given by (\ref{phistar}), after which 
the system approaches the radiation dominant universe described by 
eq.~(\ref{aasy}) as $\phi \to \phi_c$.  In order to confirm these 
behaviors, we have numerically solved the background equations 
(\ref{dilatoneq1})-(\ref{dilatoneq3}) together with the equation for 
$\beta$: 
\beq
\frac{d}{dt} S = \frac{d}{dt} \left( 
\beta^2 \frac{\partial F^{(d)}}{\partial \beta}\right) = 0\,.
\label{adiab}
\eeq
Here we assumed an adiabatic evolution corresponding to the conservation of 
the entropy $S$.  For the pure massless state (\ref{freed0}) the inverse 
temperature, $\beta=1/T$, is proportional to the scale factor $a$ as in the 
standard radiation-dominated universe.  We show in figs.~\ref{scalef} and 
\ref{varphi} one example of the bouncing solution as dotted curves for the 
pure massless state without the axion .  The dilaton continues to evolve 
toward the weakly coupled regime as expected, and the asymptotic behavior 
of the scale factor is found to be radiation dominant.

Notice that the scale factor grows from the beginning for 
positive initial values of $\dot{\mu}$.  In this case the time derivative of 
the dilaton, $\dot{\vp}=\dot{\psi}+d\dot{\mu}$, can be positive even when 
$\dot{\psi}$ is negative initially.
Since we are considering the weak coupling regime where $e^{\vp} \ll 1$
is satisfied, it is typical that the r.h.s. of eq.~(\ref{dilatoneq1}) is 
negligible relative to the first and second terms.  In the absence of the axion,
$\dot{\psi}$ is roughly estimated as $\dot{\psi} \simeq -\sqrt{d}\dot{\mu}$.
Then  $\dot{\vp}$ is positive for $d>1$, thereby leading to the growth of 
the dilaton toward the strongly coupled regime.  We argue that positive initial 
values of $\dot{\mu}$ are not so welcome in order to avoid the breakdown of 
the string-gas description at finite temperature.  In contrast, the bouncing 
solutions ({\em i.e.}, $\dot{\mu}<0$ initially) correspond to the decreasing 
dilaton toward the weakly coupled regime.

When the massive effect is taken into account, one has an extra source term 
in the small dimensions as well as the extra energy and pressure in 
eqs.~(\ref{dilatoneq1})-(\ref{dilatoneq3}).  Let us study the leading terms 
that have an explicit dependence on the radius of the small dimensions 
($r=e^{\nu}$) in the infinite sums appearing in eq.~(\ref{free}).  Taking 
the first KK and winding modes along a small direction with $p=1$, {\em 
i.e.}, the terms with $\{N_m=\bar N_m=0, m_i = (1,0,\ldots ,0), n_i=0\}$ (as 
well as $m_i$ and $n_i$ exchanged), plus the remaining $8-d$ inequivalent 
permutations, the pressure along the small dimensions is described as 
\beq
P^{(9-d)}= V_d \, C(\beta)
\beta \Biggl[\frac{1}{r^{(d+3)/2}} K_{(d-1)/2}\bigg(\frac\beta r\bigg)
-r^{(d+3)/2} K_{(d-1)/2} (\beta r) \Biggr]\,.
\label{presmall}
\eeq
Here $K_{(d-1)/2}$ is the modified Bessel function, $C(\beta)$ is defined 
as $C(\beta)=(2\pi/\beta)^{(d+1)/2} (18-2d)D(0)^2/\pi$ with $D(0)=16$ 
being a string degeneracy factor.
We have an extra energy and pressure along the large dimensions
due to the massive state in addition to (\ref{E0rad}) and (\ref{P0rad}).  See 
ref.~\cite{Bassett:2003ck} for the explicit forms of these terms.

Numerically we solved the dilaton gravity equations 
(\ref{dilatoneq1})-(\ref{dilatoneq3}) and $(\ref{adiab})$ 
together with the equation for the small dimensions 
with radius $r=e^{\nu}$: 
\beq 
\ddot{\nu}-\dot{\psi}\dot{\nu}=
\frac12 e^{\psi}P^{(9-d)}\,. 
\eeq
Since we are considering the case where the axion exists only 
for the large $d$ dimensions, we do not include its effect along the 
small dimensions. 
The pressure $P^{(9-d)}$ completely vanishes at the self-dual 
radius $r=1$ due to the compensation of KK and winding modes.  
Therefore the small dimensions are kept small around $r=1$ provided that 
the initial values of $\nu$ and $\dot{\nu}$ are close to zero.  We have done 
numerical simulations for a variety of initial conditions and found that the 
massive effect for the large dimensions is weak as long as 
the small dimensions stay nearly constant around $r=1$ (see 
figs.~\ref{scalef} and \ref{varphi} as an example).  Although the massive 
state gives rise to the extra energy and pressure, the dominant 
contribution comes from the pure massless state.
For the ideal case where the small dimensions are kept small around $r=1$,
the system is well approximated by the state of pure radiation.

\begin{figure}
\epsfxsize = 4.5in \epsffile{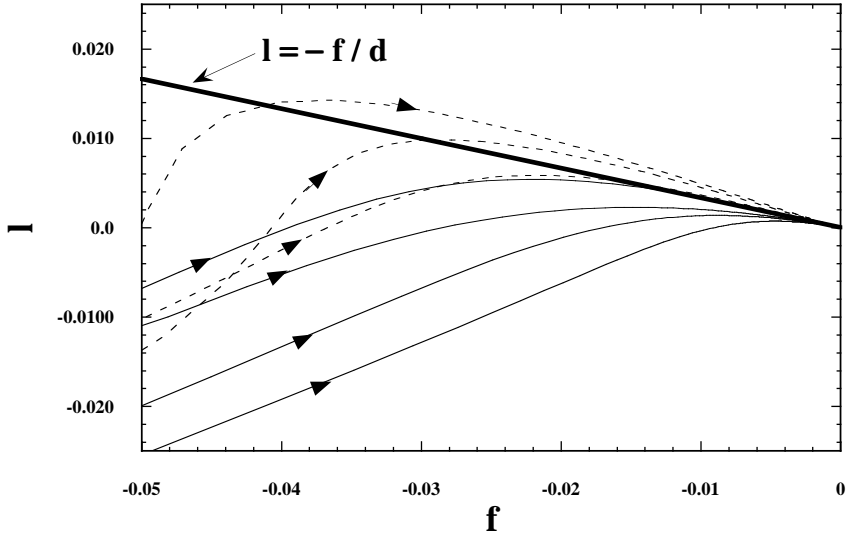}
\caption{Phase space trajectories of the solutions of string-gas 
cosmology for $d=3$.  We implement the effect of the massive state
together with the leading massless state. In the absence of the axion
(solid curves), the solutions asymptotically approach the line given by 
$l/f=-1/d$, which represents the radiation dominant universe.  The 
expansion rate $l$ gets larger when the axion is included (dotted curves), 
but the solutions eventually approach the radiation dominant universe.}
\label{phase}
\end{figure}

Eliminating the energy $E$ from eqs.~(\ref{dilatoneq1}) and 
(\ref{dilatoneq3}) and making use of the approximate relation
$E \simeq dP$ in eq.~(\ref{dilatoneq2}), we have 
\begin{eqnarray}
\label{f}
\dot{f}&=& \frac12 dl^2+\frac12 f^2+\frac12 nB^2 
\frac{e^{(n+1)\vp}}{a^{2d}} \,, \\
\label{l}
\dot{l} &\simeq& -\frac12 l^2+fl+\frac{f^2}{2d}+\frac{d(1-n)-1}{2d}
B^2\frac{e^{(n+1)\vp}}{a^{2d}}\,,
\end{eqnarray} 
where $f$ and $l$ are defined as $f \equiv \dot{\psi}$
and $\dot{\lambda} \equiv l$.
In the absence of the axion ($B=0$), it was shown in 
refs.~\cite{TV,Bassett:2003ck} that the line $l=-f/d$ is an attractor for 
the massless state.  {}From fig.~\ref{phase} we find that the solutions 
without the axion (solid curves) asymptotically approach the line $l=-f/d$ 
even in the presence of the massive state.

When the axion is included in the large $d$ dimensions, 
the evolution of the scale factor and the dilaton should be changed 
as discussed in sec.~3.1.  As one example we show in figs.~\ref{scalef} and 
\ref{varphi} the evolution of $a$ and $\vp$ for $d=3$ and $n=-1$.  The 
initial conditions are chosen to be $\dot{\vp}_i<0$ and $e^{\vp_i} \ll 1$.  
The dilaton does not continue to evolve toward the weakly coupled region 
[see eq.~(\ref{phis})] except for the initial stage, unlike the case 
without the axion.  In fact, we find from fig.~\ref{varphi} that the 
dilaton changes to evolve toward the strong coupling regime after the initial 
decrease.  The presence of the axion provides additional source terms in 
eqs.~(\ref{f}) and (\ref{l}), thereby leading to the different evolution of 
the scale factor as shown in fig.~\ref{scalef}.  When $n=-1$, the growth of 
the scale factor is faster than in the case without the axion due to the 
positive last term in the r.h.s. of eq.~(\ref{l}).  Nevertheless we find 
from fig.~\ref{phase} that the solutions eventually approach the line 
$l=-f/d$ as $f$ and $l$ decrease toward zero.  This means that the dilaton 
approaches a finite value with the radiation dominant universe ($a \propto 
t^{2/(d+1)}$ as $\xi \to \xi_c=\sqrt{M}$), as discussed in sec.~4.1.  In 
addition the scalar curvature decreases to zero as $\xi \to \xi_c$.

Since we are now dealing  with the case of $\omega=-1$ and $A=0$, 
the condition (\ref{ome1}) is satisfied whereas (\ref{omecon}) is not.  
The condition (\ref{omecon}) corresponds to the bouncing solutions where the 
scalar curvature $R$ converges in the limit of $\phi \to 0$, whereas $R$ 
diverges in the present case in this limit ($\omega=-1, A=0$).  
Nevertheless, when the universe starts out from a weak coupling regime with a 
finite curvature, it is possible to construct bouncing solutions that 
asymptotically approach the radiation dominant universe with finite
$R$ and $\vp$ irrespective of the presence of the axion as found in sec.~4.1 
for the pure massless case.  The examples plotted in figs.~\ref{scalef} and 
\ref{varphi} show the existence of nonsingular bouncing solutions 
even when the condition (\ref{omecon}) is not fulfilled.  Notice that in 
the presence of the axion but without the radiation-like source matter 
\cite{Copeland:vi} $R$ and $\vp$ diverge in both asymptotic limits.  
This indicates that inclusion of the matter source such as the ideal 
string-gas is crucial to construct the bouncing solutions without 
singularities in asymptotic future.  It will be interesting to extend our 
analysis to other matter sources which appear {\em e.g.,} in the 
$p$-brane (D$p$-brane) cosmology.

We also made numerical analysis for larger initial values of the dilaton 
which are close to $e^{\vp} \sim 1$. We found that the system tends to 
evolve toward the strongly coupled regime with the growth of $\vp$ toward 
infinity.  The scalar curvature shows divergent behavior together with 
the unbounded growth of the dilaton.  Nevertheless we have to keep in mind 
that the string-gas description based on the free energy (\ref{free}) 
ceases to be valid when the system enters the regime of strong coupling and
large curvatures.  In addition we need to consider the higher-order 
corrections to the tree-level effective action in this regime.  In the next 
section we will discuss the dynamics of the system when the dilatonic 
higher-order corrections are taken into account.

\section{Inclusion of higher-order corrections to the tree-level action}

In order to understand the effect of the higher order corrections to the 
tree-level action, let us consider the dilaton-dependent corrections, given by 
\cite{Gasperini:1996fu} 
\begin{eqnarray}
 {\cal L}_c= -\frac12 \alpha' \lambda \xi(\vp) \left[ c_1 R_{\rm GB}^2+ 
 c_2 (\nabla \vp)^4 \right]\,,
\label{lagalpha}
\end{eqnarray}
where $\xi(\vp)$ is a general function of $\vp$.  $\lambda$ is an additional 
parameter depending on the species of strings.  For example, 
we have $\lambda=-1/4, -1/8$ for the Bosonic and Heterotic string, 
respectively.  The Gauss-Bonnet term, $R_{\rm GB}^2 
=R^2-4R^{\mu\nu}R_{\mu\nu}+ R^{\mu\nu\alpha\beta}R_{\mu\nu\alpha\beta}$, 
has the property of keeping the order of the gravitational equations of 
motion unchanged.  The dilatonic corrections ${\cal L}_c$ are the sum of 
the tree-level $\alpha'$ corrections ($i=0$) and the quantum $i$-loop 
corrections ($i=1, 2, 3, \cdots$), with the function $\xi(\vp)$ 
given by \cite{Bru,Tsujikawa:2002qc,Cartier:2001gc} 
\begin{eqnarray}
\xi(\vp)=-\sum_{n=0} C_i e^{(i-1)\vp} \,,
\label{xifunction} 
\end{eqnarray}
where $C_i$ ($i \ge 1$) are the coefficients of $i$-loop corrections with 
$C_0=1$.

If the universe starts out from a sufficiently weak coupling regime ($e^{\vp} 
\ll 1$), it can reach the high curvature regime while the string coupling is 
still small. In this case the $\alpha'$ correction ($n=0$) becomes important 
before the quantum loop corrections begin to work.  Gasperini {\it et al.} 
\cite{Gasperini:1996fu} showed that inclusion of the $\alpha'$ correction 
gives rise to the fixed points in the $(\dot{\vp}, H)$ plane when only the 
dilaton is present.  This leads to an inflationary solution driven by the 
constant $H$ and the growing $\vp$.  Although the successful transition 
from this dilaton-driven inflationary phase to a decelerated FRW era is 
difficult to be achieved only in the presence of the $\alpha'$ correction, 
this problem can be solved by implementing the quantum loop corrections.  
In fact Brustein and Madden \cite{Bru} showed that inclusion of the 
quantum loop corrections can lead to the successful graceful exit due to 
the violation of the null energy condition.

Here we wish to analyze the dynamics of the system 
in the presence of the axion, modulus and the stringy matter source.
Let us consider the case of $d=3$ and $\omega=-1$.
The source terms due to ${\cal L}_c$ in eqs.~(\ref{b1})-(\ref{b3}) 
are given by \cite{Cartier:2001gc,Tsujikawa:2002qc} 
\begin{eqnarray} \label{C_i}
\rho_c=\sum_{i=0} C_i \left\{\rho_c \right\}_i\,,~~~ p_c=\sum_{i=0} C_i
\left\{p_c \right\}_i,~~~ \Delta_c=\sum_{i=0} C_i \left\{\Delta_c 
\right\}_i\,,
\label{sum}
\end{eqnarray}
where 
\begin{eqnarray}
\left\{\rho_c \right\}_i &=& \alpha' \lambda
\dot{\vp} e^{(i-1)\vp} \left\{-24c_1(i-1)H^3+3c_2\dot{\vp}^3 \right\}\,, \\
\left\{p_c \right\}_i &=& \alpha' \lambda e^{(i-1)\vp} \left\{8c_1(i-1)H 
\left[(i-1)\dot{\vp}^2H+\ddot{\vp}H+ 2\dot{\vp}(\dot{H}+H^2) \right] 
+c_2\dot{\vp}^4 \right\}\,, \\
\left\{\Delta_c \right\}_i &=& \alpha' \lambda e^{(i-1)\vp} 
\left\{24c_1(i-1)H^2(\dot{H}+H^2) -3c_2\dot{\vp}^2 
\left[4\ddot{\vp}+4\dot{\vp}H+ (i-1)\dot{\vp}^2 \right] \right\}\,.
\label{corred2}
\end{eqnarray}
Substituting these expressions for eqs.~(\ref{b1})-(\ref{b3}), we get the 
equations for $\dot{H}$ and $\ddot{\vp}$ : 
\begin{eqnarray}
\label{c1} 
& &\left[6-24\alpha'\lambda c_1 H^2 \sum_{i=0} C_i(i-1)e^{i\vp}\right] 
\dot{H}=-12H^2+\omega\dot{\vp}^2-6H\omega \dot{\vp}-2\omega 
\ddot{\vp}-\dot{\chi}^2 \\ \nonumber & & -ne^{-(n-1)\vp}\dot{\sigma}^2 
+\alpha' \lambda \sum_{i=0} C_i e^{i\vp} 
\left\{24c_1(i-1)H^4-3c_2\dot{\vp}^2 [4\ddot{\vp}+4H\dot{\vp} 
+(i-1)\dot{\vp}^2] \right\}\,, \\
\label{c2} 
& &\left[2-8\alpha'\lambda c_1 H^2 \sum_{i=0} C_i(i-1)e^{i\vp}\right]
\ddot{\vp}=-4H\dot{\vp}+4\dot{H}+6H^2+(\omega+2)\dot{\vp}^2+\dot{\chi}^2
\\ \nonumber 
& &+e^{-(n-1)\vp}\dot{\sigma}^2 
-\alpha' \lambda \sum_{i=0} C_i e^{i\vp} 
\left\{8c_1(i-1)H[(i-1)\dot{\vp}^2H+2\dot{\vp}(\dot{H}+H^2)]
+c_2 \dot{\vp}^4 \right\}+
2e^{\vp}p\,.
\label{corred}
\end{eqnarray}
In the case of $\dot{\chi}=\dot{\sigma}=0$ and $\rho=p=0$, the solutions 
approach the string phase characterized by 
constant $H$ and $\dot{\vp}$ provided that the system begins from the 
weakly coupled regime ($e^{\vp_i} \ll 1$).  These values are obtained by 
setting $\dot{H}=0$ and $\ddot{\vp}=0$ in eqs.~(\ref{c1}) and (\ref{c2}) 
with $i=0$.  The effect of the quantum loop corrections ($i \ne 0$) begin 
to work around $e^{\vp} \sim 1$, which can lead to the graceful exit toward 
the declerating FRW branch.  The number of e-folds, 
$N_e \equiv {\rm ln}\,a$, 
before the graceful exit is approximately estimated as $N_e \simeq 
|\vp_i|/2$, where $|\vp_i|$ is the initial value of the dilaton.

Let us study the case of $\omega=-1$, $\dot{\chi} \ne 0$ and 
$\dot{\sigma} \ne 0$, and $\rho=p=0$.  
In the dilaton-driven phase corresponding to the 
lowest order PBB cosmology, the evolution of $a$ and $\phi$ is described by 
$a \propto (-t)^{1/\sqrt{3}}$ and $\phi=e^{-\vp} \propto 
(-t)^{\sqrt{3}+1}$.  {}From eq.~(\ref{ana}) $\dot{\chi}$ and $\dot{\sigma}$ 
evolve as $\dot{\chi} \propto (-t)^{-1}$ and $\dot{\sigma} \propto 
(-t)^{\sqrt{3}-n(\sqrt{3}+1)}$.  Therefore $\dot{\chi}$ grows during the 
dilaton-driven phase, whereas $\dot{\sigma}$ decreases for 
$n<(3-\sqrt{3})/2$ .  The growth of $\dot{\chi}$ can change the dynamics of 
the system through eqs.~(\ref{c1}) and (\ref{c2}).  For a fixed value of 
the initial Hubble rate, the initial $\dot{\vp}$ gets larger from the 
constraint equation (\ref{b1}) in the presence of the modulus.  We also 
find from eq.~(\ref{c2}) that the increase of $\dot{\chi}$ leads to the 
faster growth of $\dot{\vp}$.  The axion has a similar effect for 
$n>(3-\sqrt{3})/2$.

\begin{figure}
\epsfxsize = 4.5in \epsffile{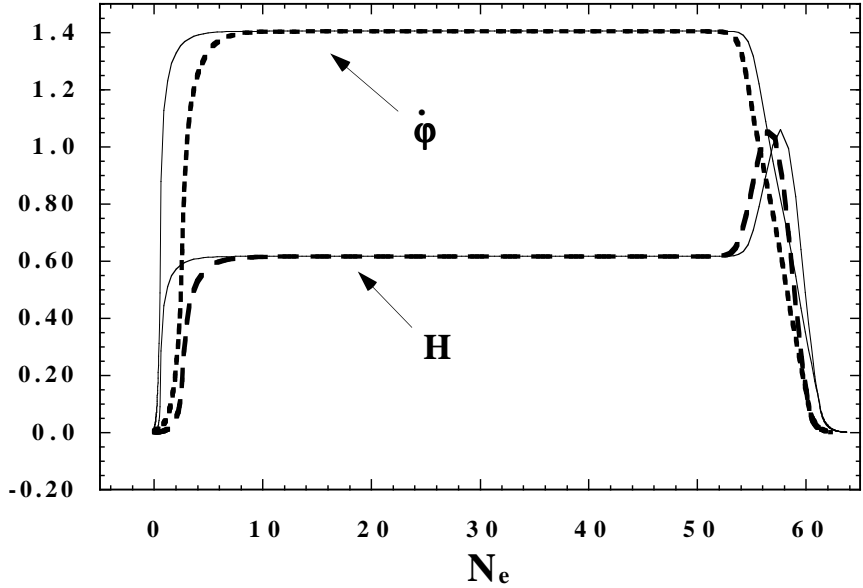} 
\caption{The evolution of $H$ and $\dot{\vp}$ as a function of the number 
of e-folds for $d=3$, $\rho=p=0$ and $n=-1$ with initial conditions 
$\vp_i=-130$ and $H_i=1.5 \times 10^{-3}$.  We include the quantum 
corrections up to $i=2$ with coefficients $C_1=1$ and $C_2=-10^{-3}$ 
together with the $\al$ correction ($C_0=1$).  The dotted curves 
correspond to the evolution without the modulus and axions fields, whereas 
these fields are involved in the solid curves with initial conditions 
$\dot{\chi}_i=10^{-2}$ and $\dot{\sigma}_i=10^{-2}$.  We do not find a 
significant difference even when $\chi$ and $\sigma$ are present.}
\label{hdpsi}
\end{figure}

\begin{figure}
\epsfxsize = 4.5in \epsffile{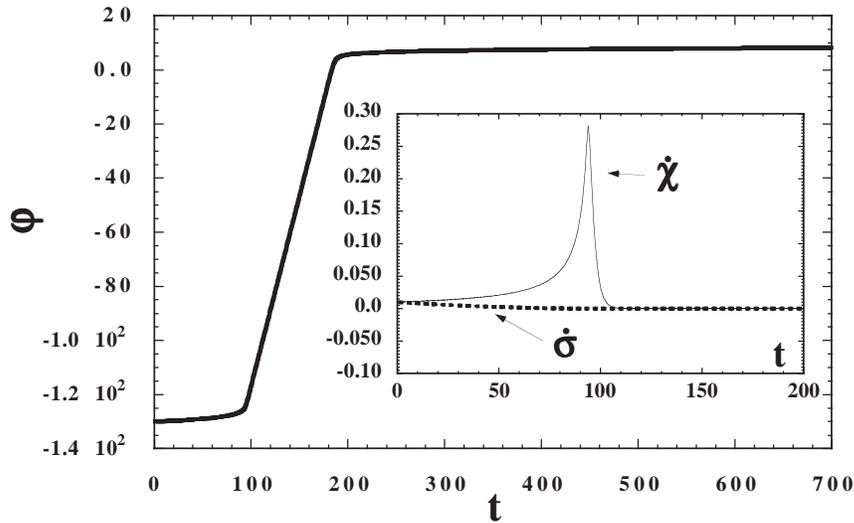} 
\caption{The evolution of the dilaton $\vp$ in the presence of
the modulus and axions fields for the same initial conditions as
in fig.~\ref{hdpsi}.  The initial dilaton-driven phase 
($0<t<100$) is followed by the string phase with a linearly growing dilaton 
($100<t<200$), after which the solutions connect to the declerating FRW 
branch.  {\bf Inset}: The evolution of $\dot{\chi}$ and $\dot{\sigma}$.
Both $\dot{\chi}$ and $\dot{\sigma}$ are exponentially suppressed
during the string phase.}
\label{vphihigh}
\end{figure}

The growth of $\dot{\chi}$ (and $\dot{\sigma}$ for $n>(3-\sqrt{3})/2$)
terminates once the system enters the string phase characterized by 
the constant $H$ and $\dot{\vp}$ ($H_0=0.616$ and $\dot{\vp}_0=1.40$ 
for $d=3$).  During this phase, we have $a \propto e^{H_0t}$ and $\vp \propto 
\dot{\vp}_0t$, thereby yielding $\dot{\chi} \propto 
e^{-(3H_0-\dot{\vp}_0)t}$ and $\dot{\sigma} \propto 
e^{-(3H_0-n\dot{\vp}_0)t}$.  The fixed point ($H_0, \vp_0$) exists in the 
range of $3H_0>\dot{\vp}_0$ in order to allow for a subsequent graceful 
exit \cite{Cartier:1999vk}.  Therefore $\dot{\chi}$ begins to decrease once 
the system enters the string phase, and the situation is the same for 
$\dot{\sigma}$ as long as $n<3H_0/\dot{\vp}_0$.  In standard low-energy 
effective string theory \cite{Copeland:vi}, we have $n=-1$ for the axion 
coupling, in which case $\dot{\sigma}$ is exponentially suppressed during 
the string phase and its energy density becomes very small relative to the 
dilaton.  Typically $n$ is smaller than unity not only in low energy 
effective string theory but also in multidimensional models with gauge fields 
or in the higher dimensional conformal gauge field theory 
\cite{Constantinidis:1999cu}.  In such cases the effect of the axion is not 
dynamically important when the correction (\ref{lagalpha}) is present.

We numerically solved the equations (\ref{c1}) and (\ref{c2}) 
for $\omega=-1$ and $n=-1$ together with the initial conditions 
satisfying $e^{\vp_i} \ll 1$, $\dot{\chi}_i<1$ and $\dot{\sigma}_i<1$.  In 
the presence of the modulus and the axion, the string phase with constant 
$H$ and $\dot{\vp}$ is typically reached earlier than in the case only with 
the dilaton.  Nevertheless the difference is not significant as seen from 
fig.~\ref{hdpsi}.  In both cases the quantum loop corrections begin to work 
around $e^{\vp} \sim 1$, after which the solutions connect to the 
declerating FRW branch.  When only the $\alpha'$ and one-loop corrections 
are present ($C_0=1$ and $C_1=1$), the dilaton and the curvature exhibit 
unbounded growth after the branch change as shown in ref.~\cite{Bru} in the 
single field case.  If the two-loop correction with negative coefficient 
$C_2$ is included in eq.~(\ref{xifunction}), this can overwhelm the 
one-loop correction when the dilaton becomes sufficiently large.  In fact, 
as shown in fig.~\ref{vphihigh}, the evolution of the dilaton becomes mild 
by including the two-loop correction.

Even when the matter source is present ($\rho \ne 0, p \ne 0$), the dynamics 
of the system is dominated by the correction (\ref{lagalpha}).  
The effect of the matter is involved in the last term of eq.~(\ref{c2}).  
Since we are considering the initial conditions corresponding to the weakly coupled 
region ($e^{\vp_i} \ll 1$), the $\alpha'$ correction tends to dominate 
over the last term in eq.~(\ref{c2}).  Then the system soon approaches the 
string phase with constant $H$ and $\dot{\vp}$.  During the string phase, 
$\rho$ and $p$ are rapidly suppressed due to the exponential growth of the 
scale factor.  This means that the matter source term is practically 
negligible around the graceful exit.  For example, the pressure of the 
string-gas discussed in the previous section evolves as $p \propto a^{-4} 
\propto e^{-2\vp}$ in the string phase, thereby leading to the exponential 
decrease of the last term of eq.~(\ref{c2}).  In fact we have done 
numerical simulations for the free energy (\ref{freed0}) in the context of 
string-gas cosmology and found that the evolution of the system is almost 
the same as shown in figs.~\ref{hdpsi} and \ref{vphihigh}.  The energy 
density and pressure of the string-gas are exponentially suppressed, 
together with the rapid decrease of the temperature ($T \propto a^{-1}$) 
before the graceful exit.

We also investigated the case where the dilatonic Brans-Dicke parameter 
$\omega$ does not equal to $-1$.  When $\omega$ lies in the range $-3/2<\omega<0$
with $d=3$ and $A=0$, the qualitative behavior is mostly the same as in the 
$\omega=-1$ case.  The difference is that the values of $H$ and $\dot{\vp}$ 
in the attractor region get bigger for smaller $|\omega|$, thereby leading 
to larger amount of e-foldings.  However, when $\omega<-3/2$, we find that 
the system is typically unstable and tends to fall into a singularity.  
Although nonsingular solutions can be obtained when $\omega$ is close to 
$-3/2$ and $e^{\vp}$ is not too much smaller than unity, the parameter 
range of these solutions is restricted to be very narrow.  This suggests 
that inclusion of the correction ${\cal L}_c$ is not necessarily welcome 
for the construction of regular solutions for $\omega<-3/2$.

\section{Summary and discussions}

In this paper we have studied the construction of nonsingular cosmological 
solutions in a general $D$-dimensional effective action derived from string 
theories.  In particular we tried to find singularity-free solutions 
analytically for the dilaton-modulus-axion system in the presence of the
``stringy'' matter source term.  We also keep the dilatonic Brans-Dicke 
parameter $\omega$ arbitrary so that the action contains a variety of 
theories such as the $F$-theory and the multidimensional theory.  In the 
tree-level action without higher-order $\al$ and loop corrections, we found 
the following results.

\begin{itemize}
\item 
Even for the general $(d+1)$-dimensional action with modulus and axion 
fields, the system is integrable for the flat FRW background if the matter 
source is radiation ($\rho=dp$) or there is no matter ($\rho=p=0$).

\item 
In the case of $\omega=-1$, we find nonsingular bouncing solutions 
where the universe starts out from the state of 
the finite scalar curvature and 
asymptotically approaches the radiation dominant stage with decreasing 
curvature and finite string coupling.  This situation naturally appears in 
the context of string-gas cosmology.

\item 
In the theories where $\omega$ is largely negative,
we have a sequence of regular bouncing solutions where the string coupling 
and the scalar curvature remain finite.  
This was already pointed out in 
ref.~\cite{Fabris:2002jt,Fabris:2002pm} for $d=3$, but we extended the 
analysis to the case of the general spatial dimension $d$ with modulus 
fields.

\item 
As a ``stringy'' matter source, we considered the ideal string-gas in thermal 
equilibrium taking into account the massive state coming from Kaluza-Klein 
and winding modes in addition to the leading massless state.
Numerically we obtained nonsingular bouncing solutions where the ``large'' 
3 dimensions asymptotically approach the almost radiation dominant universe
irrespective of the presence of the axion. 
This indicates that inclusion of the radiation-like source is crucial for the 
existence of such regular solutions.

\end{itemize}

While bouncing solutions can be obtained in the tree-level action with 
the radiation-like matter source, it can happen that the system enters the 
strongly coupled region.  In this case it is inevitable to implement the 
higher-order corrections to the tree-level action.  Therefore we also 
analyzed the case where the dilatonic higher-derivative and loop 
corrections are taken into account for the dilaton-modulus-axion system 
with some matter source.  We find that the dynamics of the system for 
$\omega=-1$ is not significantly changed by including scalar fields or 
matters other than the dilaton.  The system approaches the string phase 
characterized by constant $H$ and $\dot{\vp}$, after which the graceful 
exit is realized through the dominance of higher-order loop corrections.  
Typically the energy densities of the modulus, axion, and the matter source 
are exponentially suppressed once the dilaton-driven inflation begins 
in the string phase. 
We also find that the system tends to be unstable for 
the case of large negative $\omega$.

There are several open issues that we do not study but that deserve further 
investigation.  While we consider the flat FRW background for simplicity, 
it should be fairly easy to extend our analysis to the closed/open FRW 
background with general $D=d+1$ dimensions, as was already done for the 
case of $d=3$ without the modulus \cite{Fabris:2002pm}.  In addition it 
is worth analyzing the anisotropic background in order to check the 
generality of the singularity avoidance.  In fact a new singularity-- 
called the determinant singularity-- appears in the presence of anisotropies 
when the action contains a Gauss-Bonnet term \cite{ani}.  It is of interest 
to investigate whether the singularity solutions we found are stable by 
including small anisotropies.

The validity of our nonsingular cosmological scenarios may be investigated 
further by analyzing the power spectra of produced density perturbations.
Recent observations of WMAP \cite{Bennett:2003bz} suggest that
the spectrum of scalar perturbations is close to scale-invariant.
In the simplest tree-level pre-big-bang scenario, the curvature perturbation 
${\cal R}$ is blue-tilted with a spectral index $n \simeq 4$ \cite{BGGMV}, 
which is incompatible with observations.
This situation does not change so much even involving the higher-order 
corrections to the tree-level action 
\cite{Cartier:2001is,Tsujikawa:2001ad}.  Nevertheless it is possible to 
have nearly scale-invariant spectra of the axion perturbations depending on 
the expansion rate of the modulus field \cite{Copeland:1997ug}.  Making use 
of the ``curvaton'' mechanism \cite{curv}, there remains a possibility to 
explain the observationally supported flat spectra if the axion plays the role 
of the curvaton (see ref.~\cite{Bozza:2002ad}).  In addition, originating 
from the proposal of Ekpyrotic cosmology \cite{ekpyr}, there are a number 
of controversial argument \cite{bounceper,Tsujikawa:2002qc} about the 
choice of the gauge conditions for scalar perturbations generated in bouncing 
cosmologies.  It is certainly of interest to analyze the evolution and the 
spectra of scalar perturbations in our nonsingular bouncing models by 
taking into account the curvaton perturbations.

\vskip 25pt
\noindent
{\Large \bf Acknowledgements}
\vskip 10pt
The author thanks Bruce A. Bassett, Monica Borunda and Marco Serone for 
useful discussions and comments.  This research is financially supported from 
JSPS (No.~04942).  


\end{document}